**The hypoexponential form for the fiber-length distribution in thermoplastic composites revisited: finite initial length, cascade fracture, and its mechanistic status**

Yuichi Masubuchi
Department of Materials Physics, Nagoya University,
Nagoya 464-8603, Japan
mas@mp.pse.nagoya-u.ac.jp

**Abstract**
In a previous work [Masubuchi et al., Compos. Sci. Technol. 134, 43 (2016)], a fiber-length distribution function was proposed as the hypoexponential, or Erlang-type, convolution of two exponential waiting lengths, motivated by two independent Poisson processes for breakage and for blocking of adjacent breaks. The present paper further evaluates this hypoexponential form by introducing a finite initial length $L_0$, and by analyzing the mechanistic basis of the buckling-induced fiber breakage process. The obtained finite-$L_0$ closed-form solution was compared to a glass fiber dataset to retroactively justify the infinite-length idealization in the earlier work for such systems while providing the complete form for processes in which fragments remain comparable to $L_0$. Concerning the mechanics of fiber fragmentation, a Mellin transform analysis of the fragmentation equation showed that scale-invariant cascades yield power laws and therefore cannot generate characteristic lengths; those lengths must instead arise from scale-breaking ingredients, particularly the cutoff region and the arrested processing history. Monte Carlo simulations of a phenomenological finite-opportunity cascade showed that the hypoexponential form does not reproduce the peak of the generated distribution but captures its exponential-like tail, implying that it serves as a useful two-parameter fitting model, although not an exact generative distribution.

## 1 Introduction

For fiber-reinforced thermoplastics [1], [2], [3], the fiber length distribution governs the mechanical properties via the critical-length theory [4], in which the population above the critical transfer length has a disproportionately large influence. Due to practical difficulties in collecting sufficient samples, the average and the standard deviation of the fiber length are often used to characterize the distribution [5], although these two moments alone do not capture the skewness or the long tail. The Weibull distribution [6] is also employed when the distribution is skewed towards short fiber lengths [7]. However, these distribution functions do not reproduce some datasets in the long-tail region [8]. For instance, the length distribution reported for carbon fiber in nylon-6 composites processed by the long-fiber thermoplastic direct (LFT-D) process exhibits exponential decay, which differs from Gaussian and Weibull distributions [8].

To reproduce long-tail behavior with an exponential-like decay, the distribution function below was proposed as a two-parameter alternative to the Weibull distribution for describing fiber length data [8].

$$P_\infty(\ell;\lambda_a;\lambda_b) = \frac{e^{-\ell/\lambda_a} - e^{-\ell/\lambda_b}}{\lambda_a - \lambda_b} \tag{1}$$

Equation 1 was originally interpreted in terms of two exponential waiting lengths associated with breakage and blocking: memoryless breakage events with characteristic length $\lambda_a$ and memoryless blocking events that separate adjacent breaks with characteristic length $\lambda_b$. The function fitted the literature data for glass- and carbon-fiber nylon-6 composites comparably to Weibull and captured the long tail better in some cases [8].

However, eq 1 assumes an infinite initial fiber length. Although LFT-D is consistent with this assumption, most real processes start at a finite length $L_0$.
For processes in which the final fragments are not much shorter than $L_0$, the corrections cannot be ignored a priori, and the long-tail behavior may differ from that discussed in the previous study [8].

Apart from the effects of $L_0$, the previous study needs discussion on the following issues. First, the independence assumption underlying the two-Poisson picture is inconsistent with the fracture mechanism. If fibers break by buckling in the flow as discussed in earlier works [9], [10], [11], [12], a break redistributes stress and subsequent breaks are correlated with the stress distribution. The blocking length $\lambda_b$ was a phenomenological device that stood in for this correlation rather than a quantity derived from the mechanism; thus, the actual stochastic process generating the length distribution remained open. Second, the comparison with the Weibull distribution treated the two functions as competitors on equal footing, but they are fundamentally different. Namely, the role of the Weibull distribution in fracture statistics comes from extreme-value theory, in which it is the weakest-link limit for the fiber strength governed by the largest flaw [13]. Applying it to the fiber length invokes a justification that does not carry over, because the length is generated by a spatial breakage process and not by an extremum over flaws. The limitations of Weibull analysis, even for its proper variable, the measured single-fiber strength, have been discussed by Thomason [14]. Equation 1 was a generative guess for the length, whereas the Weibull formula is a limit law for a different variable. The previous paper [8] compared them without noting this fundamental distinction.

Since the earlier study, the modeling of fiber breakage has advanced considerably. Phelps et al. [15] tracked a discrete fiber-length distribution coupled with a mold-filling simulation, and Bechara et al. [16] proposed a phenomenological attrition model calibrated using controlled experiments in a Couette rheometer. Hohoff et al. [17] examined the effect of fiber dispersion on breakage in simple shear flow. Kang et al. [18] and Huang et al. [19] refined the buckling criterion for fibers that are constrained in thin-walled parts. Although these advances predict how the distribution evolves under given processing conditions, the functional form of the distribution and the stochastic process underlying it remain unsettled.

In this study, the abovementioned problems are addressed. First, the exact fragment-length distribution of the two-Poisson process for finite $L_0$ is derived in closed form. The result is verified against stochastic simulation and the glass-fiber data. Second, the fragmentation mechanism is analyzed using the Mellin transform and Monte Carlo methods, and eq 1 is discussed in relation to buckling-driven cascade fracture.

## 2 Fragment distribution for finite initial length

Let us recall that eq 1 is the density of a sum of two independent exponential variables with rates $\mu_a = 1/\lambda_a$ and $\mu_b = 1/\lambda_b$ (a hypoexponential, or phase-type distribution with two sequential phases):

$$P_\infty(\ell) = \int_0^\ell \mu_a e^{-\mu_a(\ell-s)} \mu_b e^{-\mu_b s} ds = \frac{\mu_a \mu_b}{\mu_a - \mu_b}\left(e^{-\mu_b \ell} - e^{-\mu_a \ell}\right) \equiv f(\ell). \qquad (2)$$

Each fragment consists of a refractory interval (mean $\lambda_b$) during which no break can occur, followed by a breakage wait (mean $\lambda_a$). On an infinite fiber, successive fragments form a renewal process with interval density $f(\ell)$; eq 1 is its interval distribution. The survival function is

$$S(\ell) = \int_{\ell}^{\infty} f(\ell')d\ell' = \frac{\mu_a e^{-\mu_b \ell} - \mu_b e^{-\mu_a \ell}}{\mu_a - \mu_b} \tag{3}$$

and the number-average length is $\bar{\ell} = \lambda_a + \lambda_b$, as before.

The model adopted here is an endpoint-initiated renewal process, in which the process starts at the fiber end and the first interval is drawn from the same density as the others. This choice differs from cutting a window out of a stationary renewal process on an infinite fiber, for which the two end intervals would follow the equilibrium residual distribution instead. The results below are exact for the endpoint-initiated model. For a fiber of finite length $L_0$, the renewal process is defined on $[0, L_0]$ with the origin as the starting point: break points occur at $0 < x_1 < x_2 < \cdots < x_N < L_0$, and each interval is drawn from $f(\ell)$. The fragments are $[0, x_1], [x_1, x_2], \cdots, [x_N, L_0]$; the last is censored by the fiber end, and if the first interval already exceeds $L_0$, the fiber is unbroken. For convenience, let us further introduce $\sigma \equiv \mu_a + \mu_b$ and $c \equiv \mu_a \mu_b / \sigma = 1/\bar{\ell}$. The renewal density, i.e., expected number of break points per unit length at distance $x$ from the origin, excluding the origin itself, follows from the Laplace transform $\tilde{f}(s) = \mu_a \mu_b / [(s + \mu_a)(s + \mu_b)]$ as

$$\tilde{u}(s) = \frac{\tilde{f}(s)}{1 - \tilde{f}(s)} = \frac{\mu_a \mu_b}{s(s + \sigma)} \tag{4}$$

Eq 4 gives the following form of $u(x)$:

$$u(x) = c(1 - e^{-\sigma x}) \tag{5}$$

Note $u(x) \to 1/\bar{\ell}$ as $x \to \infty$.

The fragment-length density has four contributions: the origin fragment, interior fragments, end-censored fragment, and unbroken fibers. The origin fragment corresponds to the first interval $[0, x_1]$ with $x_1 = \ell < L_0$, whose left endpoint is the origin. The origin is not a renewal point counted by $u$, and thus its contribution is $f(\ell)$ directly.

$$g_0(\ell) = f(\ell), \quad 0 \le \ell \le L_0 \tag{6}$$

For the interior fragments, endpoints are renewal points. A fragment of length $\ell$ whose left endpoint is a renewal point at $x \in (0, L_0 - \ell)$ contributes $u(x)f(\ell)$.

$$g_{int}(\ell) = cf(\ell)\left[(L_0 - \ell) - \frac{1}{\sigma}\left(1 - e^{-\sigma(L_0 - \ell)}\right)\right] \tag{7}$$

The end-censored fragment starts at the final renewal point $x_N = L_0 - \ell$ and survives to the fiber end with probability $S(\ell)$;

$$g_{end}(\ell) = u(L_0 - \ell)S(\ell) = c\left(1 - e^{-\sigma(L_0-\ell)}\right)S(\ell) \qquad (8)$$

The case of no interior renewal is excluded since $u(0) = 0$. Finally, if $x_1 > L_0$, the whole fiber survives as one fragment of length $L_0$ and the weight is

$$w_0 = S(L_0) \qquad (9)$$

The distribution is thus given by the following expression.

$$P(\ell|L_0) = \frac{1}{Z}[g_0(\ell) + g_{int}(\ell) + g_{end}(\ell) + S(L_0)\delta(\ell - L_0)] \qquad (10)$$

Here, $Z$ is the normalization constant given as

$$Z = \int_0^{L_0} (g_0(\ell) + g_{int}(\ell) + g_{end}(\ell))d\ell + S(L_0)$$
$$= 1 + c\left[L_0 - \frac{1 - e^{-\sigma L_0}}{\sigma}\right] \qquad (11)$$

The second expression follows because the number of fragments exceeds the number of break points by one. Thus, the normalization constant equals unity plus the expected number of break points in the fiber. The expected number of fragments produced from one initial fiber, and conservation of the total length requires the number-average fragment length to equal the initial length divided by this constant. This relation provides an independent check of the derivation.

All terms in eq 10 are elementary, and the expansion gives explicit coefficients for the basis $\{e^{-\mu_b\ell}, e^{-\mu_a\ell}, \ell e^{-\mu_b\ell}, \ell e^{-\mu_a\ell}\}$, together with two terms in which the sign of the exponent is reversed. The latter should be evaluated as $e^{-\sigma(L_0-\ell)-\mu_b\ell}$ and $e^{-\sigma(L_0-\ell)-\mu_a\ell}$ to avoid overflow, since the combined exponents are then non-positive. In the limit $\lambda_a \to \lambda_b$, the prefactors $1/(\lambda_a - \lambda_b)$ require the usual Erlang-2 degeneration, $f(\ell) \to \ell e^{-\ell/\lambda_a}/{\lambda_a}^2$.

Let us discuss the obtained form in relation to eq 1. As $L_0 \to \infty$, $w_0 \to 0$, the censoring and origin terms become $O(1)$ against the $O(L_0)$ bulk of eq 7, and $Z \to L_0/\bar{\ell}$. Consequently, the normalized distribution converges to $f(\ell) = P_\infty(\ell)$, and thus, the earlier idealization is the exact $L_0 \to \infty$ limit. The reversed-exponent terms in eq 8 act only for $\ell$ near $L_0$ and truncate the tail at the physical maximum, unlike eq 1. Finally, $\ell = L_0$ corresponds to a discrete spike of unbroken fibers.

Because endpoint conventions for renewal processes are error-prone, eq 10 is verified against a numerical simulation in which a segment of initial length $L_0$ is fragmented according to the double-Poisson process. The intervals are drawn as refractory-plus-wait sums and accumulated until they exceed $L_0$. The obtained fragments are recorded for $2 \times 10^6$ fibers. Figure 1 shows the result, with parameters cap L sub 0 over ℓ bar equal to 2.5, and finite-length effects are not negligible. The red curve drawn by eq 10 agrees well with the simulation results shown by circles; the ratio of the two over the bulk region has a standard deviation of 0.7% and a maximum local deviation of 2.6%. The unbroken fibers appear at $L_0 = 2$, and the weight is $1.35 \times 10^{-2}$ for the simulation and $1.34 \times 10^{-2}$ in eq 10. This weight is the fraction of unbroken fibers among all fragments, which equals the survival probability per initial fiber divided by the normalization constant. The survival probability itself is ~0.04 for these parameters.

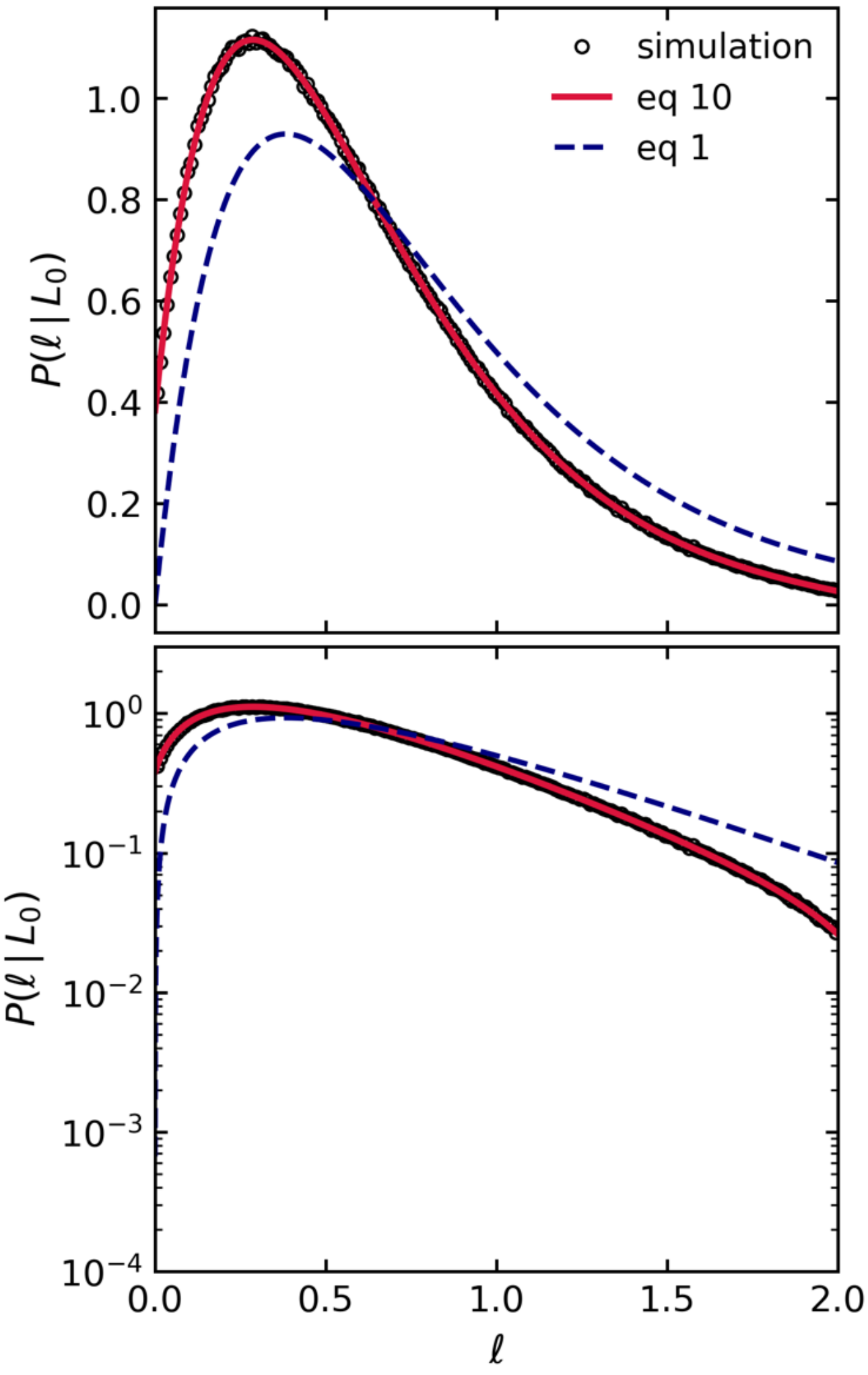

**Figure 1** Verification of the finite-$L_0$ closed form (eq 10) against direct stochastic simulation with $\lambda_a = 0.3$, $\lambda_b = 0.5$, and $L_0 = 2.0$. The circles show the simulation

result for $2 \times 10^6$ fibers. The red solid and blue dashed lines are eqs 10 and 1, respectively. Top and bottom panels display linear and semi-logarithmic plots.

Although Figure 1 clearly demonstrates that eq 1 is not capable of describing the case with $L_0 \sim \bar{\ell}$, eq 1 works reasonably well when $L_0 \gg \bar{\ell}$. Figure 2 shows a comparison of eq 10 with a dataset for glass-fiber composites [20], for which $L_0$ = 4.5 mm, whereas the fragments are of the order of 0.1–0.3 mm. Table I lists the fitting parameters $\lambda_a$ and $\lambda_b$, the survival probability per initial fiber $S(L_0)$, and the coefficient of determination $R^2$. Expanding eq 10 for $L_0/\bar{\ell} \gg 1$ gives the asymptotic form of eq 12.

$$P(l|L_0) \cong f(l)\left(1 - \frac{l}{L_0}\right) + \frac{S(l)}{L_0} \qquad (12)$$

Although this expression is useful for considering the finite-length correction, eq 1 works well for this dataset: its fitted curves are essentially indistinguishable from those of eqs 10 and 12 for this dataset (not shown). In Table II, the parameters, except for $S(L_0)$, are listed for eq 1, and the values are similar, yielding similar $R^2$. For carbon fiber composites processed via LFT-D, eqs 1 and 10 also apply, as expected, since the initial length is effectively unbounded. For mildly processed systems with nearly uniform initial lengths, eq 10 is required to represent the unbroken-fiber atom and the truncation at $L_0$, whereas eq 12 provides the leading finite-$L_0$ correction to the continuous part when $L_0/\bar{\ell}$ is large

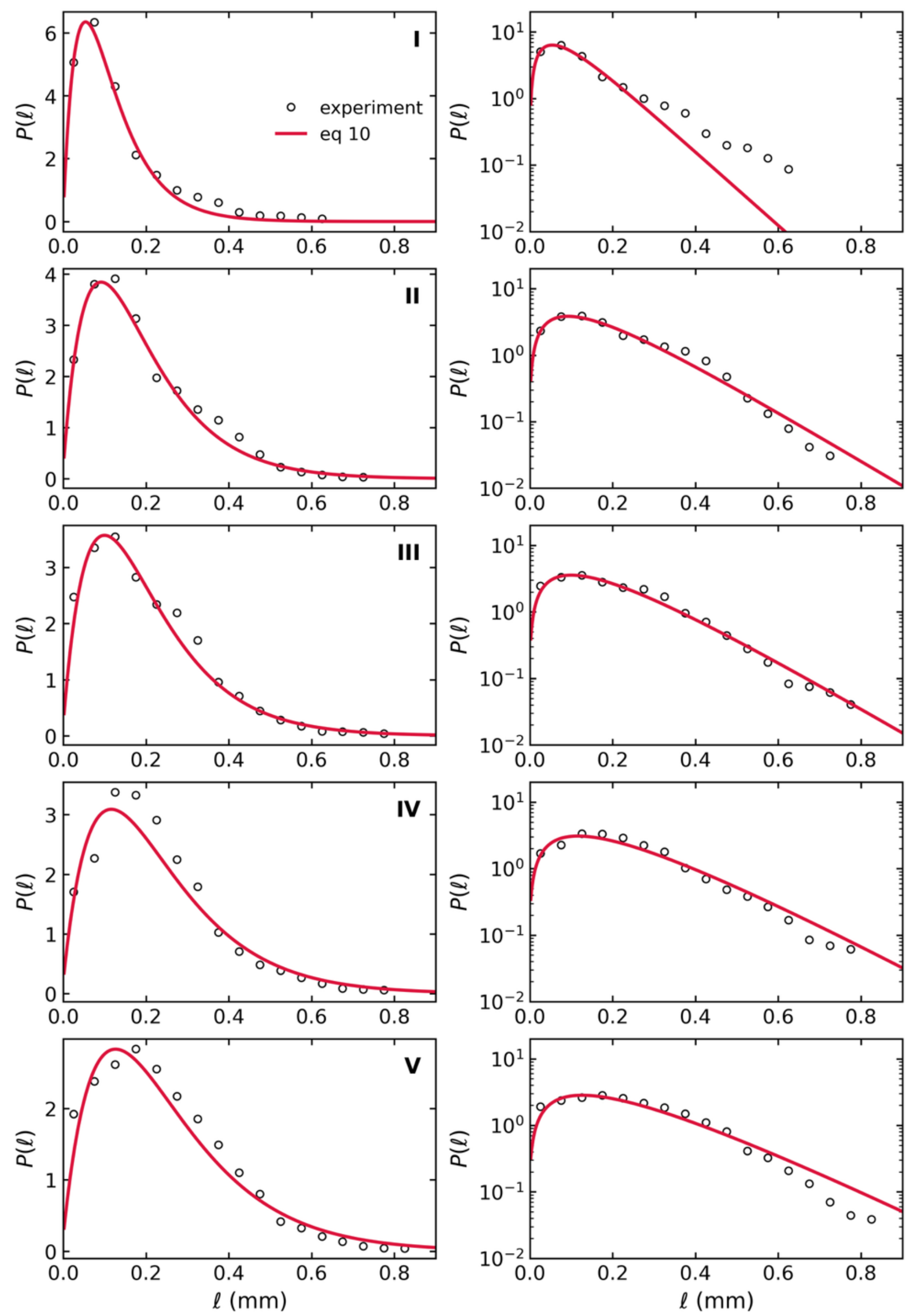

**Figure 2** Glass-fiber length distributions for materials I-V of Ularych et al. [20], from top to bottom, fitted with eq 10, with the initial length fixed at its known value $L_0$ = 4.5 mm. Circles and red curves represent the experimental data and the finite-$L_0$ theory, respectively. Left and right panels display linear and semi-logarithmic plots.

**Table I** Parameters of the finite-$L_0$ fit (eq 10) to the glass-fiber data of Ularych et al. [20], with the initial length fixed at its known value of 4.5 mm, together with the computed survival probability per initial fiber and the coefficient of determination.

| Material | $\lambda_a$ (mm) | $\lambda_b$ (mm) | $S(L_0)$ | $R^2$ |
|---|---|---|---|---|
| I | 0.040 | 0.080 | $8\times10^{-25}$ | 0.987 |
| II | 0.077 | 0.121 | $2\times10^{-16}$ | 0.984 |
| III | 0.090 | 0.122 | $4\times10^{-16}$ | 0.974 |
| IV | 0.122 | 0.122 | $4\times10^{-15}$ | 0.942 |
| V | 0.134 | 0.134 | $9\times10^{-14}$ | 0.956 |

**Table II** Parameters of the glass-fiber data shown in Figure 2 for eq 1.

| Material | $\lambda_a$ (mm) | $\lambda_b$ (mm) | $R^2$ |
|---|---|---|---|
| I | 0.038 | 0.080 | 0.987 |
| II | 0.068 | 0.125 | 0.984 |
| III | 0.072 | 0.135 | 0.970 |
| IV | 0.118 | 0.118 | 0.945 |
| V | 0.128 | 0.128 | 0.950 |

**3 Fragmentation process**

In this section, the fragmentation is considered as a stochastic process in relation to fiber buckling. Following the formulation of Meyer et al. [21], let $n(\ell, t)$ be the number density of fibers with length $\ell$ at time $t$. With a breakage rate $K(\ell)$ and a split-position kernel $\beta(x|\ell')$, which is normalized and symmetric, the evolution of $n(\ell, t)$ is written as follows.

$$\frac{\partial n(\ell, t)}{\partial t} = -K(\ell)n(\ell, t) + 2\int_{\ell}^{\infty} K(\ell')\,\beta(\ell|\ell')n(\ell', t)d\ell' \qquad (13)$$

Because pure breakage has no non-trivial steady state, and all fibers are ground toward $\ell \to 0$, a meaningful terminal distribution requires a mechanism that prevents breakage at short lengths. Buckling of fibers supplies such a mechanism, since a fiber shorter than a critical length does not buckle and does not break; then $K(\ell)$ vanishes below a certain cutoff length, $\ell_c$. This picture is behind the blocking length $\lambda_b$ in eq 1, although not mathematically exact.

Let us consider the kernel $\beta$. For buckling-driven breakage, the mode shape of a slender fiber is a function of the reduced coordinate along the fiber rather than of the absolute position, and the position of largest bending stress scales with the fiber length. The local breakage probability obtained by Durin et al. [22] from an Euler buckling analysis combined with defect statistics is likewise a function of the reduced coordinate alone. Thus, it is reasonable to assume the scale-invariant form $\beta(x \mid \ell') = \ell'^{-1}b(x/\ell')$, with $b$ defined on $(0,1)$, and let $K(\ell) = \kappa$ in the scale-invariant range above the cutoff.

To clarify the role of the Mellin transform [23], let us consider the separable power-law mode;

$$n(\ell, t) = C \exp[\omega(s)t]\, \ell^{-s} \qquad (14)$$

Substitution into eq 13 gives

$$\omega(s) = \kappa\big[2\tilde{b}(s) - 1\big], \qquad \tilde{b}(s) = \int_0^1 b\,(u) u^{s-1}\, du \qquad (15)$$

Hence $2\tilde{b}(s) = 1$ identifies a zero-growth power-law mode. This local zero mode should not be confused with a normalizable global steady-state population, because pure fragmentation has no non-trivial steady state. To formulate the cascade without this ambiguity, let us define the time-integrated breakage-event density

$$A(\ell) \equiv \int_0^\infty K\,(\ell) n(\ell, t)\, dt. \qquad (16)$$

Integration of eq 13 over time gives

$$n(\ell, \infty) - n(\ell, 0) = -A(\ell) + 2\int_\ell^\infty \beta\,(\ell \mid \ell') A(\ell')\, d\ell' \qquad (17)$$

In the scale-invariant range away from both the initial-length source and the cutoff sink, the left-hand side may be neglected, yielding the following condition.

$$A(\ell) = 2\int_\ell^\infty \frac{1}{\ell'} b\left(\frac{\ell}{\ell'}\right) A(\ell') d\ell' \qquad (18)$$

For a power-law function, $A(\ell) = \ell^{-s}$, substitution into eq 18 gives the Mellin eigenvalue condition.

$$2\tilde{b}(s) = 1, \qquad \tilde{b}(s) = \int_0^1 b(u) u^{s-1} du \qquad (19)$$

For uniform splitting, $b = 1$ and $\tilde{b}(s) = 1/s$, and the physical root is $s = 2$, which is the classical $\ell^{-2}$ fragmentation law [23]. For the end-avoiding kernel $b(u) = 6u(1-u)$, which is consistent with a centrally peaked buckling mode, $\tilde{b}(s) = 6/[(s+1)(s+2)]$. The algebraic equation formally gives $s = 2$ and $s = -5$; however, $s = -5$ lies outside the convergence domain $s > -1$ of the Mellin integral. Thus, $s = 2$ is the admissible root. The conclusion that the breakage-event density scales as $A(\ell) \propto \ell^{-2}$ is structural within the normalized, symmetric, scale-invariant class considered here and does not depend on the detailed form of $b$, since a scale-invariant cascade selects an exponent and not a length. This universality follows from the symmetry of the kernel, since the mean split position is at the fiber center for any normalized symmetric kernel, which yields the exponent 2 as a root of eq 19 in all cases.

In contrast, if the kernel is not scale invariant, the multiplicative convolution structure on which the Mellin analysis rests is lost, and the present treatment does not apply; that case is outside the scope of this paper.

The exponent obtained here is the one realized by the cascade. Equation 18 is not an independent ansatz; it follows from the time-integrated balance, eq 17, in the scale-invariant interior where the source and sink terms can be neglected. The power-law solution is the mode that transmits the flux from the source at the initial length to the sink at the cutoff, whereas the modes of eq 15 with a non-zero growth rate describe transient growth or decay of the instantaneous population and do not carry that flux.

Durin et al. [22] obtained a local breakage probability distribution of the same class, with the probability being maximal at the fiber center and vanishing at the ends. The end-avoiding character assumed here is thus supported independently. Consequently, the two characteristic lengths of eq 1 cannot arise from the self-similar regime; they must instead reflect scale-breaking features such as the cutoff region and the arrested processing history, including $\ell_c$ and the shape of $K(\ell)$ in its vicinity.

The exponent thus refers to the scaling of breakage events transmitted through the cascade, not the instantaneous number density. The distribution that accumulates below the cutoff is obtained by integrating this flux. For a sharp cutoff with uniform splitting, the terminal population below $\ell_c$ is obtained directly. Fibers above $\ell_c$ produce breakage events following the $C\ell^{-2}$ scaling obtained above, and the resulting deposition at $\ell < \ell_c$ is given below.

$$n_\infty(\ell) = 2\int_{\ell_c}^{\infty} \frac{1}{\ell'} C\ell'^{-2} d\ell' = C{\ell_c}^{-2} \tag{20}$$

The result is a constant; the terminal distribution is therefore flat, with neither a peak nor an exponential tail. For the end-avoiding kernel, the same calculation gives the following.

$$n_\infty(\ell) = 12C\left(\frac{\ell}{3{\ell_c}^3} - \frac{\ell^2}{4{\ell_c}^4}\right), \qquad 0 \le \ell \le \ell_c \tag{21}$$

The result is a polynomial with an interior peak at $\ell = 2\ell_c/3$. The end avoidance of the buckling mode thus produces the peak. However, a sharp cutoff truncates the distribution at $\ell_c$ and produces no tail. Consequently, the functional form is polynomial rather than hypoexponential.

The buckling stress is a continuous function of fiber length. Thus, the sharp cutoff assumed above, in which $K$ is constant above the cutoff, is an idealization, since $K(\ell)$ physically rises continuously above $\ell_c$. As a continuous-time rate model,

$$K(\ell) = \begin{cases} 0, & \ell \leq \ell_c, \\ \kappa[1 - \exp\{-(\ell - \ell_c)/l_w\}], & \ell > \ell_c, \end{cases} \quad (22)$$

where $\kappa$ is the rate constant and $l_w$ is a characteristic decay length. The infinite-time solution of eq 13 carries no fibers longer than the cutoff whenever the breakage rate is positive there, because the survival probability of an unbroken fiber decays exponentially in time. A distribution with a tail beyond the cutoff therefore cannot be obtained from eq 13 at infinite time.

For the continuous-time process governed by eq 13, a fragment of length $\ell$ generated at time $t_b$ and observed at the end of processing, $t_f$, has the survival probability

$$S_{\mathrm{ct}}(\ell \mid t_b, t_f) = \exp[-K(\ell)(t_f - t_b)] \quad (23)$$

and the corresponding breakage probability

$$q_{\mathrm{ct}}(\ell \mid t_b, t_f) = 1 - \exp[-K(\ell)(t_f - t_b)] \quad (24)$$

Because this probability depends on the generation time $t_b$, daughter fragments generated later in the cascade have less time available for subsequent breakage. Therefore, a finite-time solution of eq 13 cannot be represented by a single length-dependent breakage probability applied identically to every newly generated fragment.

Thus, let us introduce a separate phenomenological finite-opportunity cascade. Each newly generated fragment is tested once and breaks with probability

$$q(\ell) = \begin{cases} 0, & \ell \leq \ell_c, \\ 1 - \exp[-(\ell - \ell_c)/l_w], & \ell > \ell_c. \end{cases} \quad (25)$$

Fragments that survive this single test are recorded. This discrete construction is not a finite-time solution of eq 13, but a simplified model of fragmentation arrested by finite processing. For a fragment of fixed length subjected to $n$ independent tests, the survival factor is

$$S_n(\ell) = [1 - q(\ell)]^n = \exp[-n(\ell - \ell_c)/l_w], \qquad \ell > \ell_c \quad (26)$$

This form means that, for a fragment of fixed length, the survival factor has an effective decay width $l_w/n$. The full cascade distribution is not obtained simply by replacing $l_w$ with $l_w/n$, because additional tests also alter the production of daughter fragments.

Because the resulting cascade with $q(\ell)$ is difficult to treat analytically, Monte Carlo simulations of this discrete cascade were performed to obtain the arrested fragment density $F(\ell)$ numerically. The parameters are an initial length of 100 times the cutoff length, a cutoff width of 0.5 times the cutoff length for the soft case, and $3 \times 10^5$ initial fibers. Each fragment receives one breakage opportunity, and fragments that survive it are recorded. The results are shown in Figure 3, where the histograms use 200 equal bins and are normalized to unit area.

For the sharp cutoff the breakage probability equals unity above the cutoff, so that every longer fiber breaks with certainty and the result coincides with the infinite-time solution. The simulation agrees with eq 21 to within 0.3% of the standard deviation of the ratio, including the interior peak at $\ell = 2\ell_c/3$. However, the distribution remains truncated. In contrast, the case of the soft cutoff with end-avoiding splitting reproduces both the interior peak and an exponential-like tail beyond $\ell_c$, because fibers moderately above the cutoff survive their opportunity with a finite probability. The structure of the cascade fixes the form of this tail.

Writing the density of fragments presented for the test as $B$, the fragments that break form the density $qB$ and those that survive form the density $(1-q)B$. As an internal consistency check, the simulation reproduces these bookkeeping identities to within 0.1% and 1.2%, respectively. Far above the cutoff, where $q$ approaches unity, the breakage-event density approaches the inverse-square cascade scaling, so that the surviving density has the asymptotic form given by the product of the inverse-square power law and the exponential survival factor. The tail is therefore not a pure exponential, and this expression is a far-tail asymptotic rather than an exact result. Consistently, the decay length obtained after removing the power-law prefactor approaches the width of the cutoff as the fitting window is moved outward: the agreement is within 2% when the window starts at 1.5 times the width above the cutoff, and within 1% when it starts at three times the width, whereas including the region immediately above the cutoff degrades it to 6%.

Figure 3 also compares the simulation data with eq 1 and the Weibull distribution, which is widely used for fiber length data. Eq 1 and the Weibull distribution were fitted by least squares with two adjustable parameters, whereas eq 21 has no adjustable shape parameter once the cutoff length is given, its amplitude being fixed by the simulated fraction of fragments below the cutoff. For the sharp cutoff, eq 21 reproduces the simulation almost exactly, with $R^2 > 0.99$, whereas the values for the Weibull distribution and eq 1 are 0.79 and a negative value, respectively. For the soft cutoff, the three functions divide the range. Below the cutoff, the coefficients of determination are 0.97 for the Weibull distribution and 0.91 for eq 21, whereas eq 1 gives 0.47. Above the cutoff, in contrast, only eq 1 follows the tail. The fitted Weibull distribution decays too rapidly for the present dataset, and the deviation reaches two orders of magnitude, whereas eq 21 has no tail by construction.

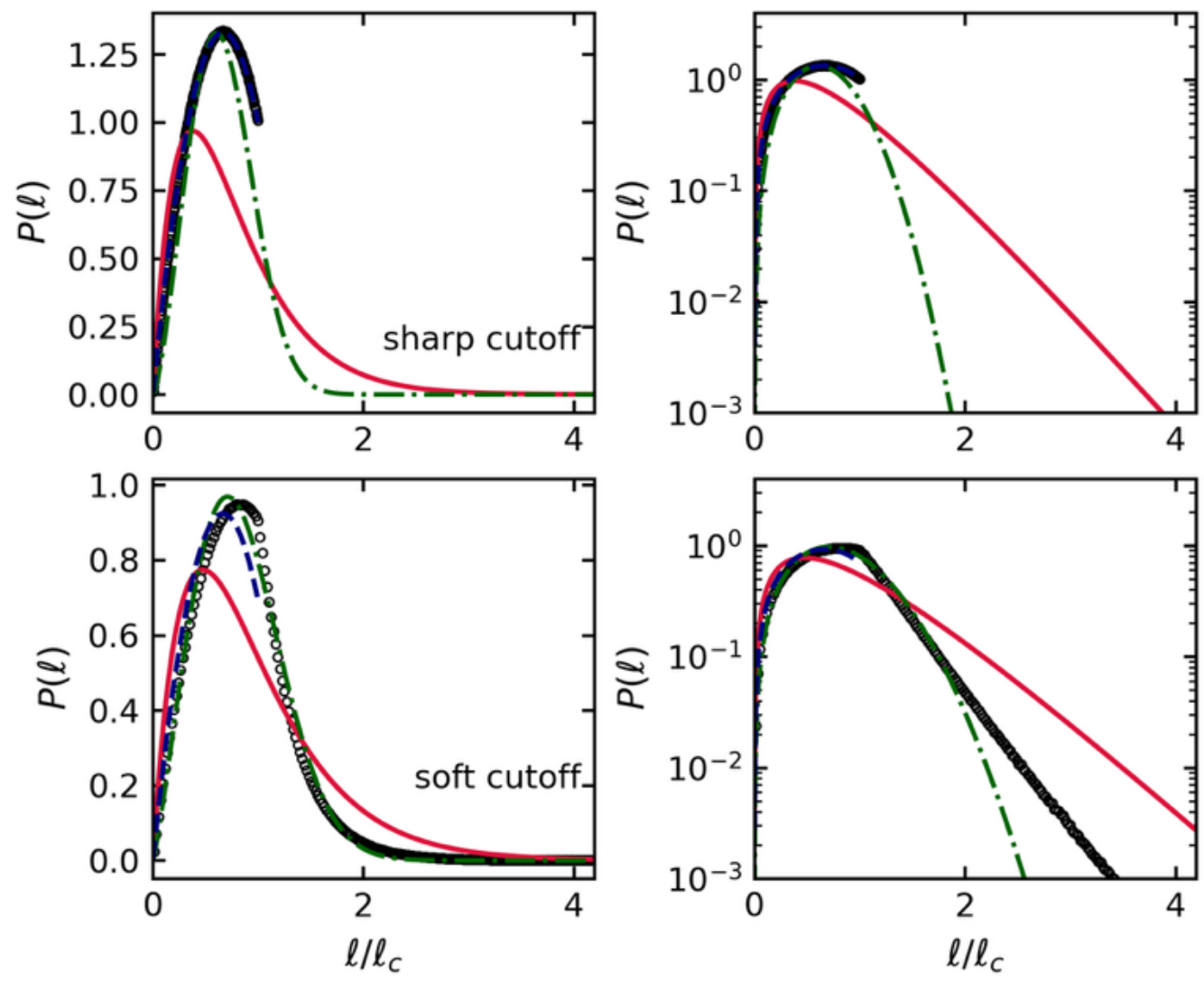

**Figure 3** Fragment distributions produced by Monte Carlo simulations of the cascades with end-avoiding splitting, for the sharp cutoff (top panels) and the soft cutoff (bottom panels). Circles show the simulation results. Red solid, green dash-dotted, and blue broken curves are eq 1, the Weibull distribution, and eq 21, respectively. Left and right panels display linear and semi-logarithmic plots.

The Monte Carlo simulation also directly verified the scaling exponent $s$ of the power-law mode. To test the scaling exponent over a wider scale-invariant range, a separate simulation with a larger initial length was performed. Recording the parent length at every breakage event in the simulation gives the breakage-event density, and a power-law fit over three to three hundred times the cutoff length yields an exponent of -2.000 for the sharp cutoff and -2.001 for the soft cutoff (data not shown). These values are consistent with $s = 2$ predicted by eqs 18 and 19.

## 4 Discussion

The results above determine the status of eq 1 as follows. The hypoexponential form is not the exact distribution generated by either the sharp-cutoff cascade or the arrested soft-cutoff cascade. The sharp-cutoff results, which coincide with the infinite-time solution, are polynomial as in eq 21, and with the soft cutoff the far tail departs from the exponential form. These deviations arise because the exact shape depends on the details of $K(\ell)$ and $\beta(x|\ell)$. For instance, a cutoff based on Euler buckling [10], [22], in which the critical stress depends on the fiber length, may improve the agreement in the tail through $K(\ell)$. Further detailed investigation using multi-fiber simulations [24], [25], [26] would be useful but is out of scope for this study. Nevertheless, as shown in Figure 3, eq 1 does not describe the peak of the simulated arrested distribution, for which the Weibull distribution and eq 21 are better. What eq 1 does capture, and the other two functions do not, is the exponential-like tail produced by the soft cutoff. Since the population above the critical transfer length can have a disproportionately large influence on the mechanical properties, eq 1 remains useful for such a purpose [27].

The parameters controlling the soft-cutoff cascade are the cutoff length, the cutoff width, and the number of breakage opportunities; only the first two are length scales. An individual assignment of $\lambda_a$ and $\lambda_b$ to distinct mechanisms cannot be made, however, because both eq 1 and eq 10 are invariant under interchange of the two lengths; the convention adopted in Table I is that the smaller value is listed first. The near-coincidence $\lambda_a \sim \lambda_b$ found for the carbon system in the earlier work is consistent with a regime in which a single effective length scale dominates the observed distribution.

Consequently, the original two-independent-Poisson derivation was not a valid mechanism, and the comparison against the Weibull distribution mixed two different kinds of objects. The proper statement is that fiber length is generated by a spatial cascade, for which eq 1, and its finite-$L_0$ completion given by eq 10, are mechanistically motivated approximations. The usual weakest-link interpretation of the Weibull distribution is an extreme-value argument for fiber strength; its use for fiber length is empirical unless a separate generative mechanism is provided.

For the evaluation of eq 10, data for mildly processed systems, in which the fragments remain comparable to $L_0$, are required to test the two features of the finite-$L_0$ form, namely the unbroken atom and the tail truncation. Controlled shear experiments by Bechara et al. [16] and by Hohoff et al. [17], in which the initial length and the applied stress are both prescribed, are possible options. Indeed, the fiber-length distribution data are available from the dead-stop experiments of Ville et al. [28] and Inceoglu et al. [29], in which samples were taken along a twin-screw extruder. However, the distributions reported by Durin et al. [22] show no unbroken-fiber atom. The apparent reason is that breakage occurs rapidly at the fiber introduction, well before the earliest sampling point.

In addition, as Sharma et al. [5] have documented for the measurement of fiber length distributions, the long-fiber end is the part most affected by sample preparation, and the atom is therefore the least likely feature to survive the measurement.

The other issue is that the initial fiber length itself has a distribution. If the initial length is drawn from a density $p_0$, the unbroken component of the unnormalized fragment-count measure underlying eq 10 becomes the following.

$$P_{unbroken}(\ell) = p_0(\ell)S(\ell) \quad (27)$$

Equation 27 is written per initial fiber and before normalization. When fragments from the whole population are pooled and counted individually, the normalization is the expected number of fragments averaged over the initial length distribution, so that eq 27 must be divided by that average. A different convention, in which one initial fiber is selected first, and then one fragment is drawn from it, would instead divide by the normalization constant evaluated at the drawn length. Equation 27 is a replica of the initial distribution, scaled by the survival probability, and is not a spike. This generalization is required for any comparison with real feedstock, for which $L_0$ is never single-valued.

The measured distributions are not the terminal state of the cascade. Durin et al. [22] set the minimum length below which a fiber cannot buckle at an aspect ratio of 2.88, which is 0.03 mm for their fibers. The peak of the measured distributions is near 0.4 mm, an order of magnitude above this cutoff. Hence, according to the analysis of the present study, the distributions are arrested transients and do not reach the final state. Consequently, the cutoff length obtained by fitting should be read as an effective arrest length, and the transition is gradual rather than sharp. This reading is consistent with the finding that the soft cutoff, rather than the sharp cutoff, reproduces the measured shape.

Interestingly, the same class of equation has been treated in a different field. Ciuperca et al. [30] analyzed a polymerization-fragmentation equation for rod-like polymers under flow, formulated on the basis of rigid-rod polymer theory, as a model for prion proliferation, and established the positivity and existence of solutions in suitable functional spaces. They used the same class of fragmentation equation as in the present study for rigid rods whose breakage is driven by flow with a length-dependent breakage rate. However, there are some substantial differences. Their model includes rod elongation due to monomer addition as well as breakage, whereas the present study is for pure fragmentation. Their breakage mechanism is based on a rigid-rod description, whereas this study considers Euler buckling. Their objective is the well-posedness of the problem, whereas the present study seeks a closed-form distribution and its asymptotic scaling. Nevertheless, the occurrence of a closely related fragmentation operator in these two physically different systems is worth noting.

## 5 Conclusions

The hypoexponential distribution function proposed in the earlier work was completed and reappraised. The fragment distribution for a finite initial length was derived in closed form and verified by simulation. Eq 10 adds an end-censoring term and an unbroken-fiber atom to eq 1, and it reduces to eq 1 as $L_0 \to \infty$. The analysis of the fragmentation process showed that the independent-Poisson motivation in the earlier work was not valid. Since scale-invariant cascades select exponents but not length scales, the characteristic lengths must arise from scale-breaking features of the cutoff region and the arrested processing history. The Monte Carlo simulations demonstrated that eq 1 does not reproduce the peak of the simulated arrested distribution, for which the Weibull distribution works better near the peak, but that eq 1 captures the exponential-like far tail obtained for a cascade arrested after a finite number of breakage opportunities. For the glass fiber data with a known initial length, the finite-length corrections were negligible. This result justifies the infinite-length idealization in the earlier work, and it also supports the use of eq 1 as a practical fitting function for $L_0 \gg \bar{\ell}$. Namely, the hypoexponential function is available as a fitting function for the long-fiber population, which can strongly influence the mechanical properties through critical-length effects, if it is read as an approximation with a known range of validity. Meanwhile, testing eq 10 against experimental data for $L_0 \sim \bar{\ell}$ remains open.

**Acknowledgments:**

**Author contributions:** The author confirms sole responsibility for the conception of the study, the derivations and simulations, the analysis of the data, and the preparation of the manuscript.

**Data availability statement:** The C source codes for the stochastic and cascade simulations, together with the data plotted in Figures 1-3, are available from the corresponding author on reasonable request.

**Conflict of interest:** The author declares no conflict of interest.

**Funding:** This study was partly supported by the Eno Science Foundation and JSPS KAKENHI (26H02291).

**Use of AI:** The author used LLMs (Claude-Opus5 and OpenAI-GPT-5.6Sol) to improve the English of the manuscript, to assist with and verify the mathematical derivations, and to draft and check the simulation and analysis codes.

**Nomenclature**

| Symbol | Meaning | Eq. |
|---|---|---|
| $\ell$ | fiber (fragment) length | - |
| $L_0$ | initial fiber length | Sec. 2 |
| $\bar{\ell}$ | number-average fragment length | 3 |
| $\lambda_a$ $\lambda_b$ | the two characteristic lengths | 1 |
| $\mu_a$ $\mu_b$ | corresponding rates, $1/\lambda$ | 2 |
| $\sigma$ | $\mu_a + \mu_b$ | Sec. 2 |
| $c$ | $\mu_a\mu_b/\sigma$ $(= 1/\bar{\ell})$ | Sec. 2 |
| $P_\infty(\ell)$ | distribution for infinite initial length | 1 |
| $f(\ell)$ | renewal interval density, equal to $P_\infty$ | 2 |
| $S(\ell)$ | survival function of the interval density | 3 |
| $u(x)$ | renewal density at distance $x$ from the fiber end | 5 |
| $g_0$, $g_{int}$, $g_{end}$ | origin, interior and end-censored fragment densities | 6-8 |
| $w_0$ | weight of the unbroken-fiber atom, $S(L_0)$ | 9 |
| $P(\ell\|L_0)$ | distribution for finite initial length | 10 |
| $Z$ | expected number of fragments per initial fiber | 11 |
| $\delta(.)$ | Dirac distribution | 10 |
| $n(\ell, t)$ | number density of fibers of length $\ell$ at time $t$ | 13 |
| $K(\ell)$ | breakage rate of a fiber of length $\ell$ | 13 |
| $\beta(x\|\ell')$ | split-position kernel | 13 |
| $b(u)$ | scale-invariant form of the split-position kernel | Sec. 3 |
| $\ell_c$ | cutoff length below which a fiber does not buckle | Sec. 3 |
| $l_w$ | width of the soft cutoff | 22 |
| $\kappa$ | rate constant of the soft cutoff | 22 |
| $\omega(s)$ | growth rate of the separable power-law mode | 14, 15 |
| $s$ | power-law exponent | 14 |
| $\tilde{b}(s)$ | Mellin transform of $b$ | 15 |
| $A(\ell)$ | time-integrated breakage-event density | 16 |
| $n_\infty(\ell)$ | fragment density deposited below the cutoff | 20, 21 |
| $q(\ell)$ | breakage probability per opportunity | 25 |
| $B(\ell),\ F(\ell)$ | densities presented for, and surviving, the test | Sec. 3 |
| $S_n(\ell)$ | survival factor after $n$ tests | 26 |
| $p_0(\ell)$ | density of the initial fiber length | 27 |
| $R^2$ | coefficient of determination | Table I, II |